\documentclass[aps,twocolumn,superscriptaddress,showpacs]{revtex4}
\usepackage{amsmath,amsthm,amssymb,amscd,float}
\usepackage{psfrag,graphicx}
\let\l\lambda  \let\be\beta  
\let\t\tau  \let\k\kappa
\def\vep{\mathcal E}
\def\ta{\tilde\alpha}
\def\tk{\tilde k}
\def\tE{\tilde E}
\def\tph{\tilde\phi}
\def\tphe{\tilde\phi_{\text{end}}}
\def\tphs{\tilde\phi_{60}} 
\def\R{\mathbf R}

\def\({{\rm(}} \def\){{\rm)}}
\newcommand{\dif}{\operatorname{d}\!}
\def\operator#1{\expandafter\def\csname#1\endcsname{\operatorname{#1}}}
\operator{sech} 
\operator{csch}
\operator{sn}
\operator{cn}
\operator{dn}
\begin{document}
\title{The Lam\'e equation in 
parametric resonance after inflation}
\author{F. Finkel}
\email{federico@ciruelo.fis.ucm.es}
\author{%
A. Gonz\'alez-L\'opez}
\email{artemio@eucmos.sim.ucm.es}
\affiliation{Departamento de F\'{\i}sica Te\'orica II, 
Facultad de Ciencias F\'{\i}sicas, 
Universidad Complutense, 
28040 Madrid, Spain}
\author{A. L. Maroto}
\email{maroto@eucmax.sim.ucm.es}
\affiliation{Departamento de F\'{\i}sica Te\'orica I, 
Facultad de Ciencias F\'{\i}sicas, 
Universidad Complutense, 
28040 Madrid, Spain}
\author{M. A. Rodr\'{\i}guez}
\email{rodrigue@eucmos.sim.ucm.es}
\affiliation{Departamento de F\'{\i}sica Te\'orica II, 
Facultad de Ciencias F\'{\i}sicas, 
Universidad Complutense, 
28040 Madrid, Spain}
%
%
\date{June 15, 2000; revised July 13, 2000}
\begin{abstract}
We show that the most general inflaton potential in Minkowski
spacetime for which the differential equation for the Fourier modes of
the matter fields reduces to Lam\'e's equation is of the form
$V(\phi)= \l\phi^4/4+K\phi^2/2+\mu/(2\phi^2)+V_0$. As an application,
we study the preheating phase after inflation for the above potential
with $K=0$ and arbitrary $\l,\mu>0$. For certain values of the
coupling constant between the inflaton and the matter fields, we
calculate the instability intervals and the characteristic exponents
in closed form.
\end{abstract}
\pacs{98.80.Cq, 98.80.Hw}
\maketitle
%
%
\section{Introduction}
The phenomenon of parametric resonance plays a fundamental role in the
modern theories of preheating after
inflation~\cite{KLS94,KLS97,STB95}. At the end of the slow-roll phase,
the inflaton field starts oscillating around the minimum of the
potential in a coherent way. If this field is coupled to some other
matter fields, the periodic evolution can give rise to an explosive
production of matter quanta thanks to the resonant amplification of
vacuum fluctuations. This production is typically characterized by the
exponential growing of the occupation number of those states whose
momentum lies within certain resonance bands. This effect is
essentially non-perturbative, and the corresponding production of
particles is much more efficient than the traditional perturbative
decay of the inflaton field during reheating. However, this
non-perturbative character makes it extremely difficult to obtain
exact results in most inflation models and, as a
consequence, one must rely on numerical computations in order to
obtain information about spectra or time evolution of the particle
production. Despite this fact, a few models are known for which
it is possible to obtain the width of the resonance bands and the
characteristic exponents in an analytical fashion. In particular, the
pure quadratic potential $V(\phi)=m^2\phi^2/2$ for the inflaton field
in Minkowski space yields the Mathieu equation for the Fourier modes
of a scalar field $\chi$ coupled to the inflaton as
$g^2\phi^2\chi^2/2$~\cite{TB90,KLS94}. In an expanding background, the
quartic inflaton potential $V(\phi)=\lambda\phi^4/4$ leads to the
Lam\'e equation for the corresponding matter fields modes~\cite{GKLS97}.
This equation also appears in connection with
some other combinations of these inflaton potentials, such as
$V(\phi)=(\phi^2-\sigma^2)^2$ \cite{Ka98}. Parametric resonance
driven by the latter potential has also been proposed 
as a mechanism for the generation of long-wavelength
pion modes from disoriented chiral condensates (DCC)~\cite{Ka99}.

The relevance of the Lam\'e equation in the study of parametric
resonance stems from its unique analytical properties, which makes it
possible to compute the resonance bands and the associated
characteristic exponents in closed form~\cite{GKLS97,Ka98}. In this
paper we extend the above results and derive the most general inflaton
potential in Minkowski spacetime for which the corresponding matter
modes equation reduces to the Lam\'e form. The resulting potential
possesses a term of the form $\phi^{-2}$, appearing typically in the
context of certain supersymmetric inflation models, see
Refs.~\cite{KR98,KR99}.

The paper is organized as follows. In Section II we present our main
result, obtaining the most general potential for the inflaton field in
a Minkowskian background leading to the Lam\'e equation for the matter
fields via the usual $g^2\phi^2\chi^2/2$ coupling. For a certain
choice of the parameters of this potential, we explicitly determine
the scaling factors necessary to reduce the equation for the Fourier
modes of the matter fields to the Lam\'e equation. In Section III, we
determine the range of values of these parameters compatible with the
slow-roll approximation and the amplitude of the density perturbations
observed by the Cosmic Background Explorer (COBE). For several values of the
coupling constant $g$ we compute (using the results of
Refs.~\cite{GKLS97,Ka98}) the resonance bands, the corresponding
characteristic exponents, and the number density of particles
produced. Our results show that the particle production in the
preheating phase remains unchanged if a term $\mu/(2\phi^2)$ is added
to the $\lambda\phi^4/4$ model for a broad range of values of the
parameter $\mu$. We finally summarize in Section IV the main results
of our paper.

\section{The inflaton potential}
Let us consider the following Lagrangian density
for the inflaton field $\phi$ coupled to a
scalar massless matter field $\chi$:
\begin{equation}
{\cal L}= \frac{1}{2}g^{\mu\nu}\partial_\mu
\phi\partial_\nu \phi+\frac{1}{2}g^{\mu\nu}\partial_\mu
\chi\partial_\nu
\chi-V(\phi)-\frac{g^2}{2}\phi^2\chi^2.
\end{equation}
We will assume our background fields to be homogeneous and
isotropic, i.e., the space-time metric is of the
Friedmann--Robertson--Walker form $ds^2=dt^2-a^2(t)\,d\mathbf x^2$, and
the inflaton field $\phi(t)$ depends only on time.
The corresponding classical equations of motion for the different
fields are
\begin{align}
&\ddot \phi+ 3H\dot \phi+V'(\phi)=0
\label{infeq}\\ &\ddot \chi + 3H\dot \chi
-\frac{1}{a^2}\partial_i\partial_i \chi + g^2\phi^2
\chi=0\,,
\end{align}
where $H=\dot a /a$, and the prime denotes differentiation with
respect to the argument of $V$. Note that we are neglecting back
reaction effects from the $\chi$ fields in the inflaton equation
\eqref{infeq} assuming that the expectation value $\langle\chi^2
\rangle$ is negligible during the first inflaton oscillations.

Let us  rescale the field as $\hat \phi=a\, \phi$, $\hat \chi=a\,
\chi$, and work in conformal $\eta$ time defined by $a\, d\eta=dt$. It
is then possible to rewrite  the equations of motion as
\begin{align}
&\frac{d^2\hat\phi}{d\eta^2} 
+a^3V'(a^{-1}\hat\phi)-\frac1a\frac{d^2a}{d\eta^2}\,\hat
\phi=0\label{eqphi}\\
&\frac{d^2\hat\chi}{d\eta^2} -
\partial_i\partial_i \hat \chi+ g^2\hat\phi^2\hat 
\chi-\frac1a\frac{d^2a}{d\eta^2}\,\hat\chi=0\,.
\label{eqchi}
\end{align}
In the particular case in which the potential is $V(\phi)=\lambda
\phi^4/4$, it can be shown~\cite{GKLS97} that the scale factor grows
as $a(\eta)\propto \eta$, and hence the last terms in
Eqs.~\eqref{eqphi}--\eqref{eqchi} vanish, while the term
$a^3V'(a^{-1}\hat\phi)$ reduces to $\lambda\, \hat \phi^3$. The
equations of motion are thus expressed in Minkowskian form in terms of
the rescaled fields. However, this is not true for an arbitrary
potential, and for that reason in the following we will concentrate
only in the Minkowskian limit ($a=1$). This can be considered as a
first approximation to the full problem and, in particular, it will
allow us to obtain analytic results in some cases. Nevertheless, we
expect that our results will carry over to an expanding universe with
minimal quantitative changes (though rather important analytic
differences) when the potential differs from $\lambda\phi^4/4$ by a
small perturbation.

Let us then consider the equations of motion in Minkowski space-time 
\begin{align}
&\frac{d^2\phi}{dt^2} + V'(\phi)=0\label{eqmphi}\\
&\frac{d^2\chi}{dt^2} -
\partial_i\partial_i  \chi + g^2\phi^2
\chi=0\,.
\label{eqmchi}
\end{align}
We shall now derive the most general inflaton potential $V(\phi)$ for 
which the differential equation for the Fourier modes of the matter 
fields
\begin{equation}
    \label{fou}
    \frac{d^2\chi_k}{dt^2}+\left(k^2+g^2\phi^2\right)\chi_k=0
\end{equation}
reduces to the Lam\'e equation
\begin{equation}
    \frac{d^2 X_k}{dx^2}+\left(\vep-mn(n+1)\sn^2 x\right) X_k = 0
    \label{lame}
\end{equation}
under a change of scale
\begin{equation}
    t =\tau x\,,\quad \phi(t) = \phi_0\,f(x)\,,
    \quad \chi_k(t)=X_k(x)\,.
    \label{scale}
\end{equation}
In the latter equations $m=\kappa^2\in(0,1)$ is the square of the
modulus of the elliptic sine function $\sn x\equiv\sn(x,\k)$.
Comparing Eqs.~\eqref{fou} and \eqref{lame} we see that we can take
without loss of generality
\begin{equation}
    f(x) = \pm\left(\be-\sn^2 x\right)^{1/2},
    \label{fdef}
\end{equation}
and
\begin{equation}
    \vep = \t^2 (k^2 +\be g^2 \phi_0^2)\,,
    \quad m n(n+1) = \t^2 g^2 \phi_0^2\,.
    \label{params}
\end{equation}
Note that $\be\ge1$ for $f$ to be real for all $x\in\R$.

It follows from Eq.~\eqref{eqmphi} that
\begin{equation}
    \frac12\left(\frac{d\phi}{dt}\right)^2+V(\phi)=E\,,
    \label{enphi}
\end{equation}
where the integration constant $E$ is the energy
density of the inflaton field.
Using the well-known identities
$$
\frac{d}{dx}\sn x = \cn x\dn x
$$
and
$$
\cn^2 x=1-\be+f^2(x)\,,\qquad \dn^2 x=1-m\be+m f^2(x)\,,
$$
it is immediate to show that
\begin{equation}
    \frac12\left(\frac{df}{dx}\right)^2 + U(f) = E_f\,,
    \label{enf}
\end{equation}
with $U$ and $E_f$ given by
\begin{align*}
U(f) &= \frac{m 
f^4}{2}+\frac12(1+m-3m\be)f^2+\frac{\be(\be-1)(1-m\be)}{2f^2}\\
E_f &= -\frac12\left[3m\be^2-2(1+m)\be+1\right].
\end{align*}
Comparison of Eqs.~\eqref{enphi} and \eqref{enf} using \eqref{scale}
shows that the inflaton potential $V(\phi)$ must be of the 
form \footnote{%
If the $\chi$ field had a mass $m_{\chi}$, the only change in the 
above derivation would be to replace $k^2$ in Eqs.~\eqref{fou} and 
\eqref{params} by $k^2+m_\chi^2$. These changes have no effect on 
Eq.~\eqref{vdef}.%
}
\begin{equation}
    V(\phi) = \l\frac{\phi^4}4 + 
    K\frac{\phi^2}2+\frac{\mu}{2\phi^2}+V_0\,,
    \label{vdef}
\end{equation}
where the potential coefficients $(\l,K,\mu)$ and the energy
density $E$ are related 
to the parameters $(\be,m,\t,\phi_0)$ through the following 
equations:
\begin{align}
    \l&=\frac{2m}{\t^2\phi_0^2}\,,\qquad
    K=\frac1{\t^2}(1+m-3m\be)\,,\label{lambdaK}\\
    \mu &=\frac{\phi_0^4}{\t^2}\,\be(\be-1)(1-m\be)\,,\label{mu2}\\
    E&=-\frac{\phi_0^2}{2\t^2}\left[3m\be^2-2(1+m)\be+1\right]+V_0\,.\label{E}
\end{align}
Note that the positivity of $\l$ forces $\mu$ to be non-negative for $V$
to have a global minimum. We have thus shown that, in order for
Eq.~\eqref{fou} to reduce under scaling to Lam\'e's equation
\eqref{lame}, the inflaton potential $V(\phi)$ must necessarily be of
the form \eqref{vdef}.

One must now prove that \emph{any} potential of the form \eqref{vdef}
leads to Lam\'e's equation \eqref{lame}. For this to be the case,
Eqs.~\eqref{lambdaK}--\eqref{E} should have a solution for
\emph{arbitrary} values of $\l>0$, $K$, $\mu\ge0$ and $E>
V_{\text{min}}$, where $V_{\text{min}}$ is the minimum value of the
potential \eqref{vdef} (the value $E=V_{\text{min}}$ must be
discarded, since then $\phi$ reduces to the trivial constant solution
$\phi=\phi_{\text{min}}$). We shall verify this only for the case
$K=0$, which we shall study in detail in the following sections.

For $K=0$  the second Eq.~\eqref{lambdaK} can be used to solve for 
$\be$ in terms of $m$, obtaining
\begin{equation}
    \be=\frac13\left(1+\frac1m\right)\,.
    \label{beta}
\end{equation}
The condition $\be\ge1$ then implies that
$$
0<m\le\frac12\,.
$$
When $0<m<1/2$ the remaining equations \eqref{mu2}--\eqref{E} are
easily solved, with the following result:
\begin{gather}
   \t^2=\frac23(\l^2\mu)^{-1/3}D(m)\label{taueq}\,,\qquad
   \phi_0^2=\left(\frac{\mu}{\l}\right)^{1/3}\frac{3m}{D(m)}\,,\\
   E = V_1\,\frac{m^2-m+1}{D^2(m)}+V_0\,,\label{Eeq}
\end{gather}
where
$$
D(m)=\left[(1+m)(2-m)\left(\frac12-m\right)\right]^{1/3},
$$
and $V_1=\frac34(\l\mu^2)^{1/3}$ is the minimum of the 
potential \eqref{vdef} for $K=V_0=0$. Since
\begin{equation}\label{tE}
\tE(m)=\frac{m^2-m+1}{D^2(m)}
\end{equation}
grows monotonically in the interval $m\in(0,1/2)$, with
$\tilde E(0)=1$ and $\lim_{m\to1/2}\tilde E(m)=+\infty$, it follows that
Eq.~\eqref{Eeq}  uniquely determines $m\in(0,1/2)$ for arbitrary values of
$E>V_1+V_0=V_{\text{min}}$.  Eq.~\eqref{taueq} then yields $\t$ and $\phi_0$ 
for arbitrary $\l,\mu>0$. For $m=1/2$ 
Eqs.~\eqref{lambdaK}--\eqref{E} simplify to
\begin{align*}
\l=\frac1{\t^2\phi_0^2}\,,\qquad \mu=0\,,\qquad
E = \frac{\phi_0^2}{4\t^2}\,.
\end{align*}
These are the equations obtained in Ref.~\cite{GKLS97} for the $\l\phi^4/4$ 
potential, which again uniquely determine $\tau$ and $\phi_0$ in 
terms of $\l>0$ and $E>V_{\text{min}}=0$. In fact, for $m=1/2$ 
Eq.~\eqref{beta} implies that $\be=1$, so that \eqref{fdef} yields 
$f^2(x)=\cn^2 (x,1/\sqrt 2)$.

\section{Exact results for the  $\lambda\,\phi^4/4+\mu/(2\phi^{2})$ 
model}

Potential terms of the form $\mu\,\phi^{-p}$ have been considered in
different contexts in the literature. They appear in hybrid inflation
models \cite{St95}, or in the so-called intermediate inflation
\cite{BL93}. But it is probably in the context of supersymmetric
models of inflation where this kind of terms is more relevant. In
fact, these contributions arise generically due to non-perturbative
effects in supersymmetric gauge theories \cite{ADS85}, the scale
$\mu$ and the exponent $p$ depending on the particular gauge group in the theory.
Inflationary models based on this kind of potentials have been studied
in Refs.~\cite{KR98,KR99}.

Motivated by the connection with the Lam\'e equation established in 
the previous section, we shall consider an
effective potential during inflation and preheating of the form
\begin{equation}
V(\phi)=V_0+\lambda \frac{\phi^4}{4}+\frac{\mu}{2\phi^2}\,.
\end{equation}
We shall derive an analytic expression for the instability intervals
and the characteristic exponents for certain values of the ratio
$g^2/\lambda$. In principle, higher-order terms could appear in the
full potential, but we will assume that their effect is negligible
during inflation and preheating. We could also have considered the
contribution of a mass term as shown in the previous section, but in
order to obtain closed expressions we will ignore it in what follows.

\begin{figure}[h]
\psfrag{t}{\begin{footnotesize}$ 10^{-6} M_P\,t$\end{footnotesize}}
\psfrag{p}[Bc][Bc][1][0]{\begin{footnotesize}$\phi/M_P$\end{footnotesize}}
\begin{flushleft}
\includegraphics[width=8cm]{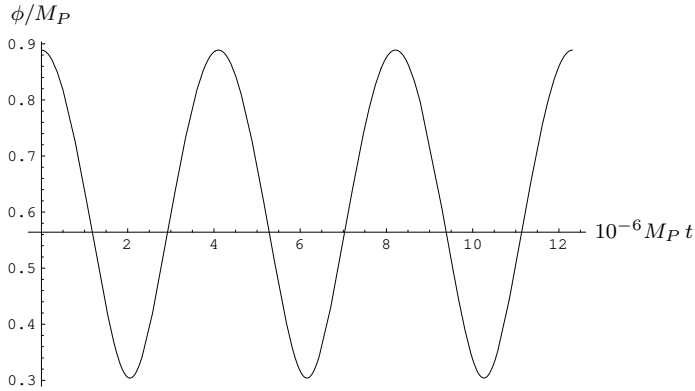}
\end{flushleft}
\begin{quote}
\caption{Oscillations after inflation of the inflaton field $\phi$ (in
units of $M_P$) as a function of time $t$ (in units of $10^6/M_p$)
computed from the exact formula \eqref{scale}--\eqref{fdef}. We have
taken $\lambda=9.091\cdot 10^{-13}$, $\mu^{1/6}=5.553\cdot 10^{-3}
M_P$, $m=0.417$, corresponding to $\ta=1$; see the
discussion in the text for details. The horizontal axis is placed at
$\phi_{\text{min}}/M_p=1/\sqrt\pi$.
\label{phiplot}}
\end{quote}
\end{figure}

The minimum of the potential is placed at
$\phi_{\text{min}}=(\mu/\lambda)^{1/6}=\alpha$, while the constant
$V_0$ must be taken as $V_0=-V_1$, i.e.~$V_{\text{min}}=0$, for the
cosmological constant to vanish. The inflaton field $\phi$ oscillates
around $\phi_{\text{min}}$ with amplitude
$\alpha\left(\sqrt{1+m}-\sqrt{1-2m}\right)/\bigl(2\sqrt{D(m)}\,\bigr)$
and period $2\tau K(\kappa)$, where $K(\kappa)$ is the complete
elliptic integral of the first kind~\cite{AS65}. Note that, in
contrast to the pure $\lambda\phi^4/4$ model, $\phi_{\text{min}}$ is
nonzero and the oscillations are not symmetric about
$\phi_{\text{min}}$ (see Fig.~\ref{phiplot}).
%
%

\begin{figure}[h]
\psfrag{a}{\begin{footnotesize}$\log_{10}\ta$\end{footnotesize}}
\psfrag{l}{\begin{footnotesize}$\log_{10}(10^{12}\,\lambda)$\end{footnotesize}}
\psfrag{m}{\begin{footnotesize}$\log_{10}(\mu^{1/6}/M_P)$\end{footnotesize}}
\includegraphics[width=8.2cm]{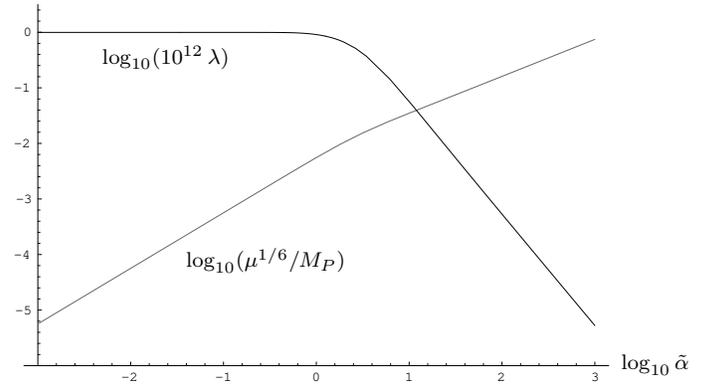}
\begin{quote}
\caption{$\log$-$\log$ plots of $10^{12}\,\lambda$
and $\mu^{1/6}/M_P$ versus $\ta$ for $10^{-3}\leq\ta\leq 10^3$.\label{figla}}
\end{quote}
\end{figure}

Let us introduce the following notation: $\tilde \phi=\phi/\alpha$
and $\tilde \alpha=\sqrt{\pi}\alpha/M_P$. The slow-roll parameters
$\epsilon$ and $\eta$ for this model are given by:
\begin{eqnarray}
\epsilon=
\frac{M_P^2}{16\pi}\left(\frac{V'}{V}\right)^2=\frac{1}{\tilde\alpha^2}
\left(\frac{\tph^4+\tph^2+1}{\tph^5+\tph^3-2\tph}\right)^2
\end{eqnarray}
and
\begin{eqnarray}
\eta=
\frac{M_P^2}{8\pi}\frac{V''}{V}=\frac{3}{2\tilde\alpha^2}\frac{\tilde
\phi^6+1}{\tilde \phi^8-3\tilde \phi^4+2\tilde \phi^2}\,.
\end{eqnarray}
The end of inflation occurs when the slow-roll approximation
breaks down, i.e., for $\epsilon \simeq 1$ or $\eta\simeq 1$. In
the case $\tilde \alpha \ll 1$ the value  $\tph_{\text{end}}$ at the end
of inflation is given to a very good approximation by
$\tph_{\text{end}}\simeq \sqrt{3/2}\,\ta^{-1}$.
However for $\ta\gg 1$ the corresponding value behaves as
$\tph_{\text{end}}\simeq 1+\frac12\,\ta^{-1}$. The initial
value $\tphs$ corresponding to 60 e-folds before inflation ends is
determined by the condition
\begin{equation}\label{phi60}
60\simeq -\frac{8\pi}{M^2_P}\int_{\phi_{\text{60}}}^{\phi_{\text{end}}}
\frac{V}{V'}\,\dif\phi\,.
\end{equation}
It follows that $\tphs\simeq \sqrt{123/2}\,\ta^{-1}$ if $\ta \ll
1$ and $\tphs\simeq 1+\frac{11}2\,\ta^{-1}$ if $\tilde\alpha \gg 1$.
Accordingly, the amplitude of the density perturbations at
$\phi=\phi_{60}$ can be written as:
\begin{align}
\delta_H(k) &=\left.\frac{16}{5}\sqrt{\frac{2\pi}{3}}\frac{V^{3/2}}{M_P^3\vert
V'\vert}\right\vert_{\phi=\phi_{60}}\label{delta}\\[1mm]
& =\frac{2}{5\pi}\sqrt{\frac{2\lambda}3}\,\ta^3
\frac{\left(\tphs^6-3\tphs^2+2\right)^{3/2}}{\tphs^6-1}\,.\notag
\end{align}
We thus get $\delta_H(k)\simeq
\frac{\sqrt{41\lambda}}{\pi}\,\left(\frac{123}5+\frac{96}{41}\,\ta^4\right)$ for 
$\tilde \alpha\ll 1$ and $\delta_H(k)\simeq \frac{242}{5\pi}\sqrt{2\lambda}\,\ta$
for $\tilde \alpha \gg 1$. The COBE normalization
$\delta_H(k)\simeq 5\cdot 10^{-5}$ determines $\lambda$ (and thus $\mu$)
as a function of $\ta$, see Fig.~\ref{figla}.
In the limit $\ta=0$ we get $\lambda \simeq 10^{-12}$, while
$\lambda\simeq 5\cdot 10^{-12}\,\ta^{-2}$ for $\tilde \alpha \gg 1$.
Since $\mu^{1/6}\lesssim M_P$ we get $\lambda\gtrsim 2\cdot 10^{-18}$
when $\tilde \alpha \gg 1$. This in turn implies
that $\tilde \alpha\lesssim 2\cdot 10^3$.

\begin{figure}[h]
    \psfrag{a}{\begin{footnotesize}$\ta$\end{footnotesize}}
    \psfrag{m}[Bc][Bc][1][0]{\begin{footnotesize}$m$\end{footnotesize}}
\includegraphics[width=8.2cm]{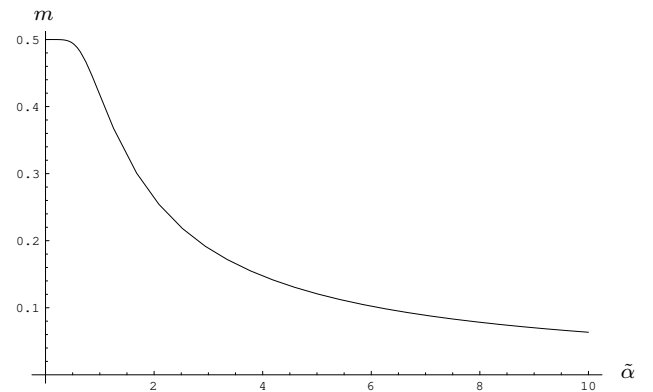}
\begin{quote}
\caption{Value of the square modulus $m$ as a function of $\ta$.\label{figm}}
\end{quote}
\end{figure}

The value of $m$ is obtained from Eq.~\eqref{Eeq}, with
the energy density of the inflaton field given by
\begin{equation}\label{Eend}
E=V(\phi_{\text{end}})=V_1\,\frac{\tphe^6-3\tphe^2+2}{3\tphe^2}\,.
\end{equation}
For $\ta\ll 1$, to a very good approximation we have
$\tE\simeq\frac34\,\ta^{-4}$. From Eq.~\eqref{tE} we thus get
$m\simeq\frac12-\frac49\,\ta^6$. On the other hand, for $\ta\gg 1$
we have $\tE\simeq1+\ta^{-2}$, leading to $m\simeq\frac23\,\ta^{-1}$.
In Fig.~\ref{figm} we present the plot of $m$ as a function of $\ta$
for $\ta$ between $0$ and $10$.\vspace*{.4cm}

If $n$ is a positive integer, the Lam\'e equation~\eqref{lame}
possesses exactly $n+1$ instability zones as the parameter $\vep$
takes values on the real line, whose corresponding solutions grow
exponentially at either $\pm\infty$~\cite{MW79}.  However, only those
instability bands for which the squared momentum $k^2$ given by
Eq.~\eqref{params} is non-negative are physically significant.  According
to Floquet's Theorem, the solutions of the Lam\'e equation in an
instability band can be written as $X_k(x)=e^{\pm\mu_k x}P_k(x)$,
where $P_k(x)$ is a periodic function and the characteristic exponent
$\mu_k$ has a nonzero real part.  The occupation number for particles with
momentum $k$ produced in the preheating phase can be estimated in terms
of the characteristic exponent as $N_k(t)\sim e^{2\mu_k t/\tau}$~\cite{Ka98}. 
Remarkably, for integer $n$ the characteristic exponent for the
Lam\'e equation is given by an exact formula involving a
quadrature which, at least for the lowest values of $n$,
can be expressed in terms of elliptic integrals~\cite{GKLS97,Ka98}. We shall
omit here most details and quote from the above references
only the main steps for the derivation of this formula.

The key ingredient for obtaining the formula is the construction of an
exact expression for the product of two linearly independent solutions
of the Lam\'e equation \footnote{See also the forthcoming paper by J.
Garc\'\i a-Bellido for an extension of this method.}. The product of
two such solutions satisfies the third-order differential equation
given by
\begin{align}
& 2p(z)M'''(z)+3p'(z)M''(z)+\Big[ p''(z)+2\big( \vep\notag\\
&\quad +mn(n+1)(z-1)\big)\Big]M'(z)+mn(n+1)M(z)=0\,,\label{eqM}
\end{align}
where $z=\cn^2 x$, and
$$
p(z)=(1-m)z+(2m-1)z^2-m z^3\,.
$$
If $n$ is a nonnegative integer, Eq.~\eqref{eqM} is satisfied by a
suitable polynomial of degree $n$, which (following~\cite{Ka98})
shall be written as
\begin{equation}\label{solM}
M_{(n)}(z)=\sum_{i=0}^n a_i^{(n)}z^{n-i}\,,
\end{equation}
with the normalization condition $a_0^{(n)}=1$. Since $z$ is a periodic function
of $x$, the polynomial $M_{(n)}(z)$ coincides with the product of two linearly
independent solutions of the Lam\'e equation~\eqref{lame} when $\vep$ belongs
to an instability zone. It was shown in Refs.~\cite{GKLS97,Ka98} that the
characteristic exponent is determined by the definite integral
\begin{equation}\label{muk1}
\mu_k=\frac{C_{(n)}}{2K(\k)}\int_0^1\frac{\dif z}{\sqrt{p(z)}M_{(n)}(z)}\,,
\end{equation}
where
\begin{equation}\label{C}
C_{(n)}^2=\big(mn(n+1)-\vep\big)(a_n^{(n)})^2+(m-1)a_{n-1}^{(n)}a_{n}^{(n)}\,.
\end{equation}
The instability intervals can be obtained imposing
that the right-hand side of Eq.~\eqref{C} be positive. The sign of
$C_{(n)}$ is chosen so that the real part of $\mu_k$ is
positive if $\vep$ lies in an instability zone.

If all the roots of the polynomial $M_{(n)}(z)$ are real and
different, the definite integral in Eq.~\eqref{muk1} can be expressed
in terms of elliptic integrals. Indeed, let $z=1-y$, and let the
constants $\be_i$, $D_i$, $i=1,\dots,n$, be defined by
\begin{equation}\label{beD}
\frac1{M_{(n)}(1-y)}=\sum_{i=1}^n \frac{D_i}{1-\be_i^{-1}y}\,.
\end{equation}
The formula~\eqref{muk1} for the characteristic exponent $\mu_k$ then
reduces to
\begin{equation}\label{muk2}
\mu_k=\frac{C_{(n)}}{K(\k)}\sum_{i=1}^n D_i\hat\Pi(\be_i^{-1}|\k)\,,
\end{equation}
with
\begin{equation}\label{hPi}
\hat\Pi(\be_i^{-1}|\k)=
\begin{cases}
\Pi(\be_i^{-1}|\k) & \quad\text{if } \be_i^{-1}<1\,,\\
K(\k)-\Pi(\be_i\k^2|\k) & \quad\text{if } \be_i^{-1}>1\,,
\end{cases}
\end{equation}
where $\Pi(s|\k)$ is the complete elliptic integral of the third 
kind~\cite{AS65}.

We shall now compute the characteristic exponent for $n=1,2,3$ using
Eq.~\eqref{muk2}. It shall be convenient to define the dimensionless
momentum $\tk=k/M_{P}$. Using~Eq.~\eqref{taueq} and the
definition of $\ta$, Eq.~\eqref{params} becomes
\begin{equation}\label{tildek}
    \vep=\frac 13\left(\frac{2\pi 
    D(m)}{\lambda\ta^2}\,\tk^2+(m+1)n(n+1)
    \right)\,.
\end{equation}
For $n=1$, the Lam\'e equation~\eqref{lame} possesses two instability
zones, namely $\vep\in(-\infty,m)$ and $\vep\in(1,1+m)$; see for
instance Ref.~\cite{FGR00}. The first instability zone is not
admissible, since it would force the momentum $k$ to take imaginary
values. Using Eq.~\eqref{tildek}, the second instability zone leads to
the resonance condition
\begin{equation}
    \label{band1}
    \frac{\lambda\ta^2(1-2m)}{2\pi D(m)}< \tk^2
    <\frac{\lambda\ta^2(m+1)}{2\pi D(m)}\,.
\end{equation}
Thus for $m\neq\frac 12$ (i.e., for $\ta\neq 0$), there is an initial
threshold for the resonant values of the momentum. If the $\chi$ field
had a mass $m_\chi$, the instability interval \eqref{band1} would be
shifted by an amount $-m_\chi^2/M_P^2$. For $m_\chi$ large enough,
this could result in the disappearance of the threshold, or even of the
whole interval.

\begin{figure}[h]
    \psfrag{a}{\begin{footnotesize}$\ta$\end{footnotesize}}
    \psfrag{m}{\begin{footnotesize}$\mu_k$\end{footnotesize}}
    \psfrag{k}{\begin{footnotesize}$10^{12}\tk^2$\end{footnotesize}}
\includegraphics[width=8cm]{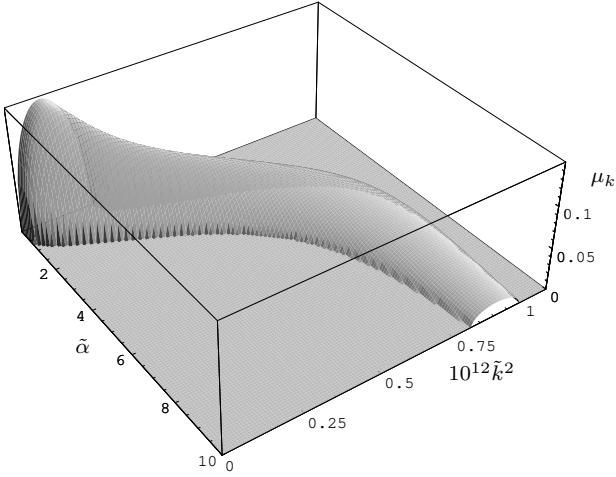}
\begin{quote}
\caption{Characteristic exponent as a function of $\ta$ and $10^{12}\tk^2$ for 
$n=1$.\label{figmu1}}
\end{quote}
\end{figure}

\begin{figure}[h]
    \psfrag{a}{\begin{footnotesize}$\ta$\end{footnotesize}}
    \psfrag{N}[Bc][Bc][1][0]{\begin{footnotesize}$\log_{10} N(\ta)$\end{footnotesize}}
    \psfrag{t}{\begin{scriptsize}$n=3$\end{scriptsize}}
    \psfrag{d}{\begin{scriptsize}$n=2$\end{scriptsize}}
    \psfrag{u}{\begin{scriptsize}$n=1$\end{scriptsize}}
    \includegraphics[width=8.2cm]{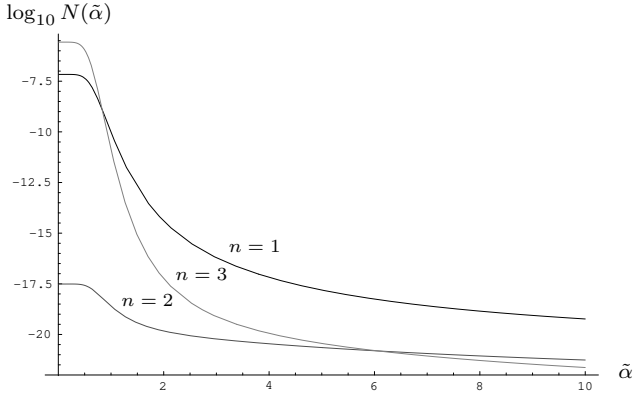}
\begin{quote}
\caption{Plot of $\log_{10} N(\ta)$
as a function of $\ta$ for $n=1,2,3$.\label{fignumden}}
\end{quote}
\end{figure}

\begin{figure}[h]
    \psfrag{a}{\begin{footnotesize}$\ta$\end{footnotesize}}
    \psfrag{N}[Bc][Bc][1][0]{\begin{footnotesize}$N(\ta)/N(0)$\end{footnotesize}}
    \psfrag{t}[Br][Br][1][0]{\begin{scriptsize}$n=3$\end{scriptsize}}
    \psfrag{d}{\begin{scriptsize}$n=2$\end{scriptsize}}
    \psfrag{u}{\begin{scriptsize}$n=1$\end{scriptsize}}
\includegraphics[width=8.2cm]{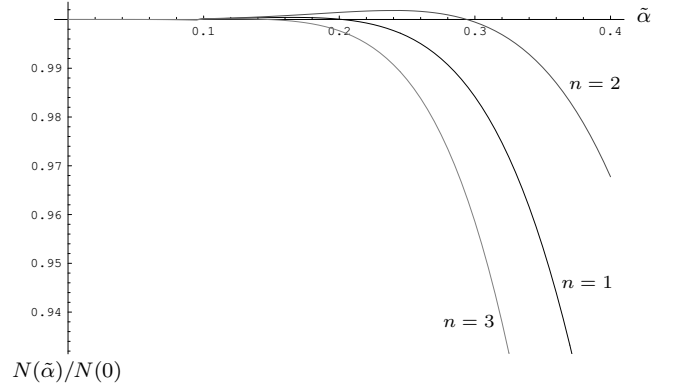}
\begin{quote}
\caption{Number densities ratio $N(\ta)/N(0)$ for small
values of $\ta$ and $n=1,2,3$.\label{figprod}}
\end{quote}
\end{figure}

The polynomial
$M_{(1)}(z)$ is given by
$$
M_{(1)}(z)=z+\frac{1-\vep}{m}\,.
$$
Therefore
$$
\be_1^{-1}=D_1=\frac{m}{1+m-\vep}\,,
$$
and
$$
C_{(1)}^2=\frac1{m^2}(\vep-m)(\vep-1)(1+m-\vep)\,.
$$
Note that $\be_1^{-1}>1$ in the second instability zone. In Fig.~\ref{figmu1}
we plot the characteristic exponent as a function of $\ta$ and $10^{12}\tk^2$
for $\ta$ between $0$ and $10$.
The absolute maximum of the characteristic
exponent for $n=1$ is $\mu_k=0.147$ at $\ta=0$ and $\tk^2=1.08\cdot 
10^{-13}$,
in agreement with~\cite{Ka98}.
The maximum of the characteristic exponent for fixed $\ta$ decreases
monotonically with $\ta$. In contrast, the width of the resonance band
of the squared momentum $\tk^2$
increases for $\ta$ small, reaching its maximum at $\ta=2.442$.
In order to compare the efficiency of the particle production
with the pure $\lambda\phi^4/4$ model, we estimate the
number density in units of $M_{P}^3$ after 30
oscillations of the inflaton field (when the back reaction becomes
significant in the $\lambda\phi^4$ model) as
\begin{equation}
    N(\ta)\simeq\frac1{2\pi^2}\int_I
    \tk^2 e^{120\mu_{k}K(\k)}\dif\tk\,,
\end{equation}
where $I$ is the union of all the instability intervals.
In Fig.~\ref{fignumden} we plot the logarithm of the number 
density versus $\ta$. In Fig.~\ref{figprod} we represent the ratios
$N(\ta)/N(0)$ for small values of $\ta$.
The number density for $n=1$ (i.e., $g^2=\lambda$) is
almost constant for $\ta\lesssim 0.2$.
Therefore, in this case the effect of the
term $\mu/(2\phi^2)$ in the energy transfer from the inflaton field
is negligible provided $\mu^{1/6}\lesssim 10^{-3}M_{P}$. The number
density decreases sharply for $\ta\gtrsim 0.5$.

We next examine the case $n=2$. In this case the Lam\'e equation~\eqref{lame}
possesses three instability zones, out of which the only relevant one
is $4+m<\vep<2\big(1+m+\sqrt{m^2-m+1}\big)$, leading to
\begin{equation}
    \frac{3\lambda\ta^2(2-m)}{2\pi D(m)}< \tk^2
    <\frac{3\lambda\ta^2\sqrt{m^2-m+1}}{\pi D(m)}\,.
\end{equation}
The width of the resonance band of the squared momentum $\tk^2$ also increases
in this case for small $\ta$, with its maximum located at $\ta=0.858$.
The coefficients $a_1^{(2)}$ and $a_2^{(2)}$ defining the polynomial
$M_{(2)}(z)$ are given by (omitting the superscript $(2)$)
\begin{equation}
a_1=\frac{4-2m-\vep}{3m}\,,\quad
a_2=\frac{(1+m-\vep)(4+m-\vep)}{9m^2}\,.\label{a1a2}
\end{equation}
One can immediately show that the roots of $M_{(2)}(z)$ are real and 
different if $\vep$ lies in the above instability band.
The coefficients $\be_{1,2}$ and $D_{1,2}$ are respectively given by
\begin{gather}
\be_{1,2}=\frac12\big(2+a_1\mp\sqrt{a_1^2-4a_2}\big)\,,\\
D_{1,2}=\frac{\pm(2+a_1)+\sqrt{a_1^2-4a_2}}{2(1+a_1+a_2)\sqrt{a_1^2-4a_2}}\,,
\end{gather}
with $a_{1,2}$ given by~\eqref{a1a2}. Note that $\be_{1,2}^{-1}>1$
in the instability zone. The coefficient $C_{(2)}^2$ is given by
\begin{align*}
C_{(2)}^2 &=\frac1{81m^4}\left(1+m-\vep\right)
       \left(4+m-\vep\right)
       \left(1+4m-\vep\right)\\
&\quad\times\left({\vep^2}-4(1+m)\vep+12m\right)\,.
\end{align*}
The maximum of the characteristic exponent in the
instability band also decreases monotonically with $\ta$.
The absolute maximum is $\mu_{k}=0.036$ at $\ta=0$,
$\tk^2=7.67\cdot 10^{-13}$, four times smaller than
the absolute maximum for $n=1$. Correspondingly, the particle 
production is also much less efficient than in the case $n=1$, see
Fig.~\ref{fignumden}. The maximum particle production for $n=2$ occurs
when $\ta=0.245$, see Fig.~\ref{figprod}.

The case $n=3$ presents some unexpected effects. In this case, the 
Lam\'e equation~\eqref{lame} possesses four instability zones, out of 
which the only relevant ones are 
$\vep\in\big(4(1+m),2+5m+2\sqrt{4m^2-m+1}\big)$
and $\vep\in\big(5+2m+2\sqrt{m^2-m+4},5(m+1)+2\sqrt{4m^2-7m+4}\big)$. 
In terms of the dimensionless momentum $\tk$, these are
\begin{equation}
    0<\tk^2<\frac{3\lambda\ta^2}{2\pi 
    D(m)}\,\big(m-2+2\sqrt{4m^2-m+1}\big)\,,\label{band31}
\end{equation}
and
\begin{align}
    & \frac{3\lambda\ta^2}{2\pi 
    D(m)}\,\big(1-2m+2\sqrt{m^2-m+4}\big)<\tk^2\notag\\
    & \qquad<\frac{3\lambda\ta^2}{2\pi 
    D(m)}\,\big(1+m+2\sqrt{4m^2-7m+4}\big)\,.\label{band32}
\end{align}
For both resonance bands, the width increases for small $\ta$,
with the respective maxima located at $\ta=1.052$ and $\ta=0.533$.
The coefficients of the polynomial $M_{(3)}(z)$ defined in Eq.~\eqref{solM}
are given by
\begin{align}
a_{1} &=\frac{9-6m-\vep}{5m}\,,\notag\\
a_{2} &=\frac{2\vep^2+(4m-26)\vep+27m^2-51m+72}{75m^2}\,,\label{a1a2a3}\\
a_{3} &=-\frac1{225m^3}\,\Big(\vep^3-2(4m+7)\vep^2+(16m(m+5)+49)\vep\notag\\
      &\qquad-12(m+1)(8m+3)\Big)\,.\notag
\end{align}
The coefficients $C_{(3)}^2$ and $\be_{i}$, $D_{i}$, $i=1,2,3$,
are then easily obtained from Eqs.~\eqref{C} and~\eqref{beD}. The
resulting expressions are very cumbersome and shall not be displayed
here. It may be shown that the coefficients $\be_{i}$ are all real
and different in the resonance bands.

Just as in the previous cases $n=1,2$, the maximum of the
characteristic exponent decreases monotonically with $\ta$ for both
resonance bands. For the lower resonance band given in
Eq.~\eqref{band31}, the absolute maximum is $\mu_{k}=0.159$ at
$\ta=0$, $\tk^2=2.09\cdot 10^{-13}$, while for the higher one in
Eq.~\eqref{band32} the maximum value is $\mu_{k}=0.0078$ at $\ta=0$,
$\tk^2=1.857\cdot 10^{-12}$. The particle production at $\ta=0$ is two
orders of magnitude more efficient than in the case $n=1$. However, it
decreases with $\ta$ much faster than in the previous cases $n=1,2$,
see Figs.~\ref{fignumden}, \ref{figprod}.

\section{Conclusions}

In this paper we have characterized the most general scalar potential
for the inflaton field leading to the Lam\'e equation for the matter
field modes in a Minkowskian background. The resulting potential
possesses a term of the form $\mu/(2\phi^2)$ in addition to the terms
$\lambda\phi^4/4+K\phi^2/2$ already studied in the literature. We have
analyzed the effect of this new term in the preheating era after
inflation in the particular case $K=0$. Exact expressions for the
resonance bands and the characteristic exponents have been derived for
certain values of the coupling constant between the inflaton and the
matter fields. The effect of the new term in the particle production
is virtually negligible provided $\mu^{1/6}\lesssim10^{-3}M_P$, even
though the inflaton potential is modified in an essential way near the
origin. However, for $\mu^{1/6}\gg 10^{-3}M_P$ matter production
is heavily suppressed by the new term as compared to the pure
$\lambda\phi^4/4$ model.
The situation in this respect is expected to remain unchanged
for other values of the coupling constant.


\begin{acknowledgments}
    A. L. M. wishes to thank J. Garc\'\i a-Bellido for useful 
    discussions.
    This work was partially supported by grants
    DGES PB98-0821 and DGICYT AEN97-1693.
    \vspace*{.5cm}
\end{acknowledgments}

\end{document}